# Evidence for Adsorbate-Enhanced Field Emission from Carbon Nanotube Fibers

P.T. Murray[1], T.C. Back[1], M.M. Cahay[2], S.B. Fairchild[3], B. Maruyama[3],
N. P. Lockwood[4], and M. Pasquali[5]

[1]Research Institute, University of Dayton, Dayton, OH 45469-0170 USA

[2]Spintronics and Vacuum Nanoelectronics Laboratory, University of Cincinnati, Cincinnati Ohio 45221 USA

[3]Materials and Manufacturing Directorate, Air Force Research Laboratory, WPAFB, OH 45433 USA

[4]Directed Energy Directorate, Air Force Research Laboratory, Kirtland AFB, NM 87117 USA

[5]Department of Chemical and Biomolecular Engineering and Department of Chemistry, The Smalley Institute for Nanoscale Science and Technology, Rice University, Houston, Texas 77251-1892 USA

**Abstract**

We used residual gas analysis (RGA) to identify the species desorbed during field emission (FE) from a carbon nanotube (CNT) fiber. The RGA data show a sharp threshold for $H_2$ desorption at an external field strength that coincides with a breakpoint in the FE data. A comprehensive model for the gradual transition of FE from adsorbate-enhanced CNTs at low bias to FE from CNTs with reduced $H_2$ adsorbate coverage at high bias is developed which accounts for the gradual desorption of the $H_2$ adsorbates, alignment of the CNTs at the fiber tip, and importance of self-heating effects with applied bias.

Field Emission (FE) sources are of interest because of their high brightness, small spot size, as well as their portable and relatively simple nature. Carbon nanotube (CNT) cathodes in FE sources have received considerable attention because of the field enhancement brought about by their high aspect ratio, resulting in the need for lower applied voltages to achieve similar emission currents. CNT cathodes have numerous potential applications that include use in flat panel displays [1], electron beam lithography [2], x-ray generation [3-5], electron microscopy, and ion propulsion [6].

Three operational regimes have been identified by others [7] for FE from CNTs. The first (regime I) occurs near the FE threshold and is one in which emission is enhanced by the presence of adsorbates initially present on the CNT surface. Subsequent studies, involving thermal field emission electron spectroscopy [8], field emission microscopy from SWNT caps [9], and by current saturation measurements [10] from adsorbate-covered SWNTs, were consistent with this assertion. Comparison of the FE electron energy distributions acquired from clean and adsorbate-covered SWNTs led [11] to the conclusion that enhancement occurred through resonant states below the Fermi level. Transition to the second regime (regime II) is accompanied by an apparent saturation in the FE current, as noted by a decrease in the Fowler-Nordheim (FN) current extrapolated from the lower field data, a change in the field emission microscopy patterns, and increasing fluctuations of the current [12]. The third (regime III) is one in which the adsorbates are removed and is characterized as intrinsic emission from bare CNTs. Similar adsorption-related effects on FE have been reported for multiwall CNTs [13-18]. Finally, enhanced emission has been reported [19] from CNTs subjected to hydrogen plasma treatment, indicating the benefit of adsorbed hydrogen on the FE properties of CNTs.

Even though this model is consistent with numerous experimental and theoretical studies, no studies have tested it directly by detecting and identifying the species desorbed from CNT fiber cathodes during FE. Here we address this issue and show that the threshold for desorption of $H_2$ occurs at a field strength that coincides with the transition from regimes I to II, thereby directly validating the model.

The cathode for this work consisted of a highly electrically and thermally conductive CNT fiber spun from a liquid crystalline dope consisting of single, double, and triple-walled CNTs dissolved in chlorosulfonic acid (CSA) [20-22]. The internal structure of the fibers consists of CNT fibrils that are approximately 10-100 nm in diameter and are held together by van der Waals forces [20-22]. The alignment and packing density of these fibrils affects the fiber's electrical and thermal conductivity. Each fibril is composed of closely packed CNTs [20-22].

FE from the (20 μm diameter) CNT fiber was carried out in an ultrahigh vacuum chamber whose base pressure was 4.0 x $10^{-7}$ Pa. A copper anode probe tip (1 mm diameter) was aligned with the CNT fiber cathode with the use of two orthogonally situated cameras looking through two different windows on the chamber. One camera was equipped with a long working distance objective that was used to accurately determine the anode-cathode gap distance. The gap was adjusted with integrated stepper motors capable of 2.5 μm steps. Once the gap distance was set, the voltage on the anode was increased, with a Keithley 6517A source meter, at a rate of 1 V per 10 s to the maximum of 1000V. Data were recorded at each voltage setting under LabView control. The residual gas analyzer (RGA) was situated in a line-of sight with respect to the FE regime, and the distance from the FE regime to the entrance aperture of the RGA was 15 cm. The RGA and field emission scans were acquired such that the intensity for each particle mass was recorded at every voltage step. The anode-cathode tip separation was 1 mm for the work presented here.

Adsorbate-enhanced FE can be manifested as hysteresis in the I-V curves [18,23,24] which is illustrated in Figure 1(a). These data were acquired first with increasing, and then with decreasing field strength; the arrows in Figure 1(a) indicate the direction of the field sweep for each curve. It can be seen that the FE current measured during the former is larger (by a factor of ~3) compared to that measured during the downward sweep. The hysteresis can be understood by assuming that adsorbates originally present on the CNT fiber are eventually removed at the highest field strengths (by field assisted and thermally-induced desorption), and that the current from the nascent, bare CNTs during the subsequent downward sweep is therefore not enhanced by adsorbed species and is lower (at the same field) than for the forward voltage sweep. In this paper, we focus on phenomena taking place with increasing field strength and provide a comprehensive FE model for the gradual transition from regimes I to III.

Figure 1(b) shows the same I-V data (for increasing field strength) presented in a semi-log plot. Three distinct regimes (I, II, and III) are clearly discernible, together with breakpoints (represented by the dashed lines) at external electric field strengths ($E_{ext}$) of 0.49 and 0.7 V/µm. The inset presents the data from regimes I and II using a FN plot; the excellent fit (denoted by the solid red line) within regime I suggests that FE from adsorbate-covered CNTs is well described by the FN formalism. The breakpoint at 0.49 V/µm suggests a change in the FE dynamics. Interestingly, near the breakpoint, in addition to exhibiting a change in slope, the data also show increased current fluctuations very similar to that reported earlier [11,12]; these fluctuations eventually decrease at higher field strength. Beyond the breakpoint at $E_{ext} = 0.7$ V/μm, the data represent intrinsic FE from bare CNTs.

Shown in Figure 2 are the RGA data. The left and right ordinates represent the change in partial pressures of $H_2$ and CO, respectively, from their base line values. The data indicate no detectable change in partial pressure of either species during FE within regime I. However, the data does exhibit a threshold for desorption of both species at a field strength of $E_{ext} = 0.49$ V/μm that coincides with the breakpoint seen in the I-V curves. These data, in conjunction with the hysteresis in the I-V curves, confirm the assumption that the breakpoint between regimes I and II correlates with the threshold for desorption. Interestingly, the change in CO partial pressure was over an order of magnitude smaller than that of $H_2$. We also detected an increase in the partial pressure of $CO_2$ (not shown) that was approximately two orders of magnitude smaller than that of $H_2$.

Next, we explain the I-V data based on a model consistent with the observation that $H_2$ was the dominant desorbed species observed in the RGA experiments. Since the fiber is composed of fibrils containing closely packed CNTs, we assume for simplicity in the following analysis that emission arises from individual CNTs. If FE is taking place and the contribution from thermionic emission is negligible, the current-voltage characteristics should be well fitted by the simplified form of the FN expression

$$\text{I (Amps)} = A_{eff} \frac{1.54 \; 10^{-6}}{\varphi} \beta^2 E_{ext}^2 \; e^{-\frac{6.83 \times 10^7 \; \varphi^{3/2}}{\beta E_{ext}}} \qquad (1)$$

where $A_{eff}$ is the effective area (in cm$^2$) of the fiber which is emitting, $\varphi$ is the work function (in eV) of the emitting surface, $\beta$ is the field enhancement factor and $E_{ext}$ is the externally applied electric field (in Volts/cm), i.e., $E_{ext} = \frac{V}{d}$, where V is the applied bias (in Volts) between anode and cathode (assumed to be same as between tip of the fiber and the anode since the potential drop along the fiber is negligible) and d is the distance (in cm) between the anode and the base of the CNT fiber cathode. In regimes I and III, the data are characteristic of FN emission and are well fitted by the following expressions

$$\ln(I_a/V^2) = 1.28 - \frac{14233}{V} \qquad (2)$$

and

$$\ln(I_i/V^2) = -17.80 - \frac{2825}{V} \qquad (3)$$

where $I_a$ represents the emission current (in A) from the adsorbate-covered CNTs, and $I_i$ represents the current (in A) from intrinsic (bare) CNTs. We found that in the intermediate region (regime II), the I-V curve is well fitted by the following expression

$$I = f_a I_a + f_i I_i \qquad (4)$$

where the functions $f_a$ and $f_i$ are factors explicitly given by $f_a = \frac{1}{1+e^{((V-539)/21.7)}}$ and $f_i = \frac{1}{1+e^{((677-V)/27.3)}}$.

We interpret the prefactor $f_a$ as the fraction of CNT emitters whose FE is due to the presence of adsorbates on their tips and sidewalls. However, we recognize that the system is more complex than this; some CNTs have adsorbed O, thereby reducing emission, and other defective or broken tip CNTs will in many cases emit better than capped CNTs. CSA decomposes into $H_2SO_4$ and HCl during coagulation and washing with water [20]. Most likely, some Cl- and S-compounds (potentially HCl, $H_2SO_4$, $SO_3$, $SO_2$, etc) remain adsorbed on the fiber. These have the two-fold effect of doping the fiber (leading to its higher electrical conductivity [20]) and giving rise to a large number of surface dipoles (H-S, H-SO, H-Cl, …) which help to lower the work function of the CNTs. We refer to these surface dipoles as X-H dipoles hereafter. We see no evidence of desorbed Cl in the RGA data suggesting it has a higher adsorption potential than $H_2$. However, a weak signal that is potentially S signal (at a mass of 32 AMU) follows the same trend as that of desorbed $H_2$, CO, and $CO_2$. Because this mass corresponds to that of atomic S as well as the diatomic $O_2$ (formed by dissociative ionization of desorbed $CO_2$ in the RGA ionizer), we cannot unambiguously state that we have detected desorbed S.

At low electric field, a majority of the CNTs are not initially aligned with the field. As the field increases past the threshold field $E_a$ (~ 0.539 V/μm), two mechanisms lead to a decrease of the prefactor $f_a$ with applied bias. First, as the external electric field increases, the charge transfer between the H atoms and the X species on the CNT surface is opposed by the external field, weakening the X-H dipoles and making the adsorbed species become more mobile on the surface of the CNTs. As two H atoms recombine and desorb (which is commonly called Langmuir-Hinshelwood desorption[31]), they desorb as a molecular species, as observed in the RGA data. Desorption of the adsorbates reduces the number of X-H dipoles on the sidewalls and near the CNT tips. The second desorption process is due to the self-heating effect near the tip of the CNT [26-28] which leads to the progressive disappearance of the X-H dipoles via thermal desorption. For a gas and surface in equilibrium, the ratio of the gas number density ($N_g$) to the two dimensional surface gas density ($N_a$) is given by [32]

$$\frac{N_g}{N_a} = Q_{rot}\sqrt{\frac{2\pi mkT}{h^2}}e^{-D_a/kT} \qquad (5)$$

where $D_a$ is the adsorption potential, $Q_{rot}$ is the rotational partition function, $m$ is the mass of the species and $T$ is the temperature of the surface and gas. The assumption of thermal equilibrium and a single species is violated, but Eq.(5) approximately describes the relationship between the gas phase and adsorbed species. With increasing electric field, there is an increase in power dissipation near the tip. This eventually leads to the removal of a majority of the X-H dipoles consistent with Eq.(5) and a concomitant increase of the work function to the value associated with a bare CNT. In regime II, there is also a progressive increase in the number of CNTs aligning with the externally applied electric field. This phenomenon is modeled by the prefactor $f_i$, which represents the fraction of CNTs with FE characteristic of bare nanotubes with a work function around 4.8eV. We interpret the threshold electric field $E_i$ (~ 0.677 V/μm) for regime III as the field strength at which the majority of CNTs are fully aligned with the applied external field.

At some critical point within regime II (at which $f_a I_a = f_i I_i$ and which is denoted $E_c$ in Figure 3), a sufficiently large fraction of the X-H surface dipoles have been removed, and emission from a single X-H dipole dominates the emission process. Below $E_c$, the effective emission area is much larger that the fiber cross-sectional area. It involves numerous low work function X-H dipoles covering the CNT tips and sidewalls. An interpretation of F-N fit of the I-V curve above $E_c$ is that emission area is much smaller because the individual CNTs line up at the higher biases resulting in a few atoms near the tips of the mostly bare CNTs contributing primarily to the observed FE current and shielding the CNT fibers below them. The individual CNTs will have a higher temperature due to the high current density they are producing and hence should have a highly reduced number of X-H dipoles compared to the bulk CNT fiber. The individual CNTs at the tip are exposed due to the mechanical cutting process of the CNT fibers which leaves individual CNTs at the surface of the fibers which are observed in Scanning Electron Microscope (SEM) images. Assuming a range of work function between 4.8 and 4.0 eV for the adsorbed species [25] and a work function of 4.8 eV for the bare CNTs, the difference in the FN fits in regimes I and III can be attributed to a difference in field enhancement factors of $\beta_i/\beta_a$ ranging from ~ 5 to 6.6. Hence, from Eq.(1), the ratio of the effective area of the fiber leading to FN emission in regimes I and III is estimated to be roughly $A_{eff,i}/A_{eff,a} \sim 10^{-10}$.

This large difference between the effective emission areas is possibly due to desorption of the X-H dipoles with increasing external field and self-heating effects in the CNTs. The latter are responsible for various physical mechanisms responsible for initiating CNT vacuum breakdown. This includes thermal runaway at large applied electric field which causes a maximum temperature at the tip of some CNTs and resulting in their gradual shortening [29]. Another possible explanation for the small effective area ratio could be due CNT fragmentation at high emission currents as a temperature leading to CNT breakdown is reached inside the CNTs away from their tips leading to a smaller number of emission sites [30].

An additional support for the above arguments comes from a closer look at the noise in the experimental data in regime II. Figure 3 shows that there is substantial noise below the critical field $E_c$, where emission mostly comes from adsorbates which are in large concentration. The emission from individual emitters depends strongly on the energy of the X-H dipoles and the exact shape of the nanotube tips. The wide variety in the tip characteristics in addition to the dynamic nature of the adsorbate coverage on the surface accounts for the large noise in the I-V data. Above $E_c$, the main contribution to FE comes from CNTs with low adsorbate coverage which is more stable in the applied electric field, and thus the noise in the I-V curve is much less.

Even though not shown here, we have found that the signal to noise ratio in the I-V data increased above $E_c$.

Far into regime III, self-heating effects in the bare CNTs start to prevail. In fact, by reducing the anode to cathode distance at a fixed applied bias of 1000V, we have observed an increase in flickering, flashing spots close to the tip of other similar fibers, followed by a sudden bright glow after which the fibers stopped emitting. This was attributed to a degradation of the emitting surface of the fiber due to runaway self-heating effects.

In summary, we have studied desorption from a CNT fiber during FE using RGA. The RGA data show the existence of a sharp threshold for desorption of $H_2$ from the fiber at a value that coincides with a breakpoint in the I-V curves, thereby confirming an earlier assumption that FE from CNTs is enhanced by the presence of adsorbates [7]. This agrees well with a comparison of the relative emission areas of the CNT fiber at low and high applied bias. This leads to the conclusion that FE is enhanced by X-H dipoles originally present on the CNT surface at low bias, and mostly bare CNTs account for the majority of FE at larger bias when most of the $H_2$ is desorbed from the surface. This conclusion is further supported by the reduction in noise in the I-V data above the critical external field strength $E_c$, where emission from bare CNTs in the fiber dominates. The I-V data can be well fitted using a FN based model which includes switching between surface dipole enhanced FE at low applied bias to FE from bare CNTs at large applied bias as desorption of the adsorbates increases with increasing CNT fiber tip temperature.

## Acknowledgments

We thank N. Behabtu, C. C. Young, and D. E. Tsentalovich for providing the CNT fibers. This work was supported by Air Force contract FA8650-11-D-5401 at the Materials & Manufacturing Directorate (AFRL/RXAP) and by AFOSR grant FA9550-09-1-0590. The authors thank John Luginsland and Joycelyn Harris at AFOSR for supporting this work.

**Figure Captions**

**Figure 1**. (a) Current-field plot from CNT fiber FE showing measured current with increasing and decreasing field sweeps and (b) same data presented in (a) during the increasing field sweep and showing three operational regimes. The dashed lines represent the breakpoints between the regimes. Inset: Fowler-Nordheim plot spanning regimes I and II.

**Figure 2**. RGA showing increase in partial pressure of (a) $H_2$ and (b) CO during field emission. The dashed line represents the transition between regimes I and II and represents the onset for desorption.

**Figure 3.** Fit (red curve) to the experimental I-E data in the intermediate regime II shown in Fig.1 using Eq.(4). The individual contributions from the nanotubes with adsorbates ($f_a I_a$) and bare nanotubes ($f_i I_i$) are shown. The noise in the data is larger to the left of the field $E_c$ where $f_a I_a = f_i I_i$.

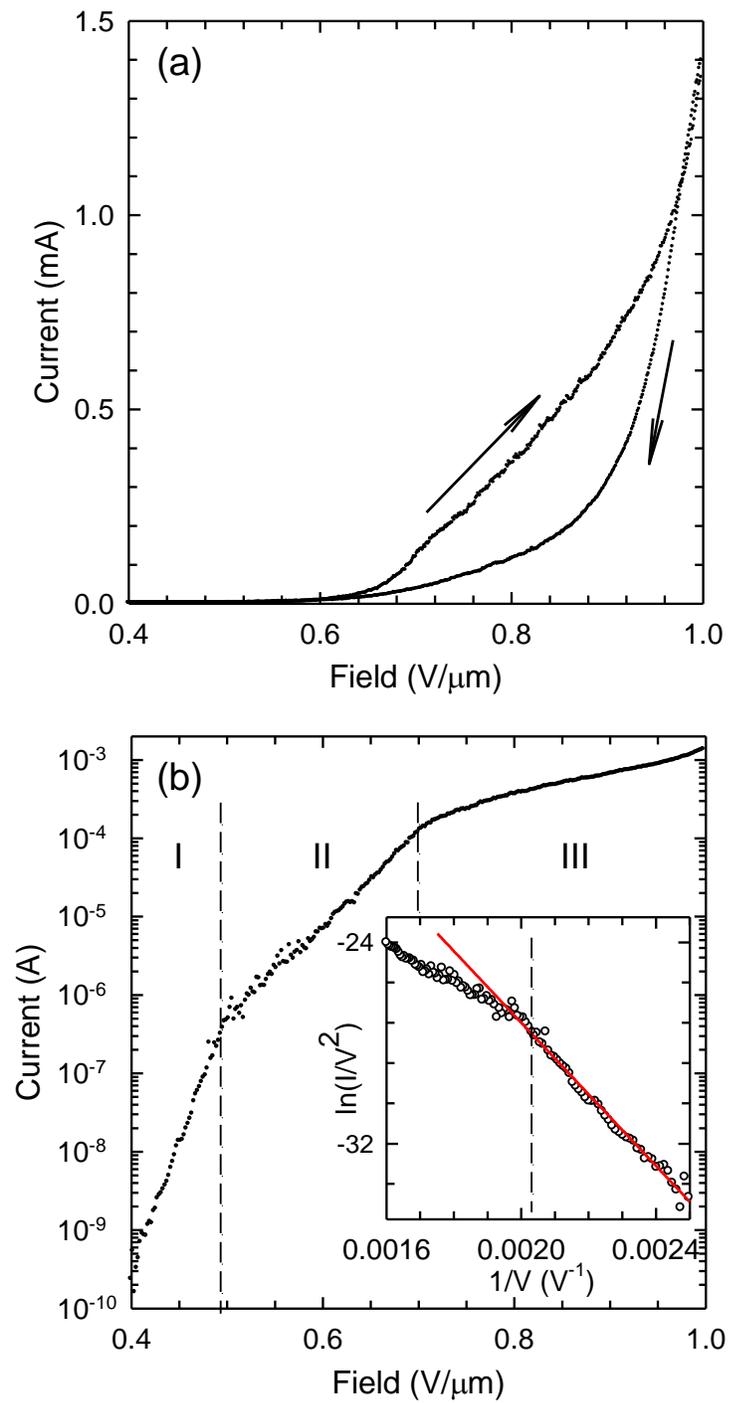

**Figure 1 (P.T. Murray et al.)**

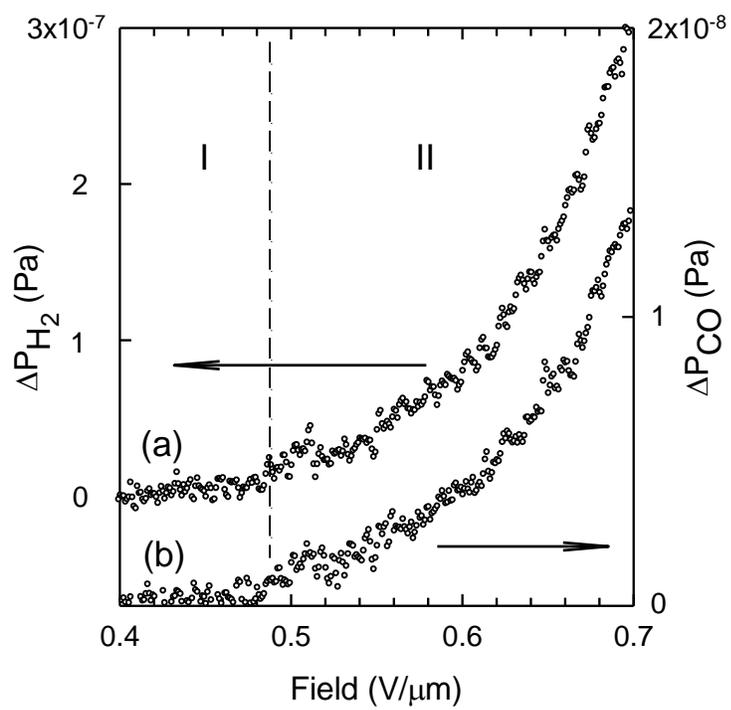

**Figure 2 (P.T. Murray et al.)**

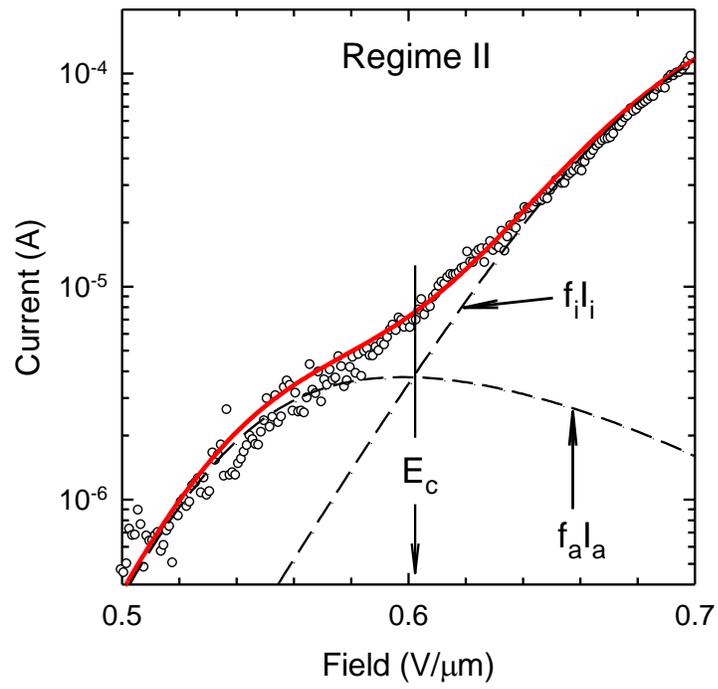

**Figure 3 (P.T. Murray et al.)**